\documentclass[prd,preprint,superscriptaddress]{revtex4}
\usepackage{revsymb}
\usepackage{amsmath}
\newcommand{\be}{\begin{equation}}
\newcommand{\ee}{\end{equation}}
\newcommand{\ben}{\begin{eqnarray}}
\newcommand{\een}{\end{eqnarray}}

\begin{document}
\title{Holographic dark energy in Brans-Dicke theory}
\date{\today}
\author{Narayan Banerjee\footnote{E-mail address: narayan@juphys.ernet.in}}
\affiliation{Relativity and Cosmology Research Centre, Department
of Physics, Jadavpur University, Calcutta -700032, India}
\author{Diego Pav\'{o}n\footnote{E-mail address: diego.pavon@uab.es}}
\affiliation{Departamento de F\'{\i}sica, Facultad de Ciencias,
Universidad Aut\'{o}noma de Barcelona, 08193 Bellaterra
(Barcelona), Spain}

\begin{abstract}
In this Letter it is shown that when the holographic dark energy
is combined with the Brans-Dicke field equations the transition
from decelerated to accelerated expansion of the Universe can be
more easily accounted for than when resort to the Einstein field
equations is made. Likewise, the coincidence problem of late
cosmic acceleration gets more readily softened.
\end{abstract}
\maketitle
\section{Introduction}
Arguably, the finding that the Universe is currently accelerating
its expansion constitutes the most intriguing discovery in
observational cosmology of recent years \cite{accelerating,wmap3}.
The long-lived Einstein-de Sitter cosmological model is no longer
fit to explain the present state of affairs, and must be replaced
by some other model compatible with a transition from decelerated
to accelerated expansion. Very often, to achieve this transition,
a novel energy component (dubbed ``dark energy"), that violates
the strong energy condition and clusters only at the largest
accessible scales, is invoked. But, aside from these two features,
nothing is known for certain about the nature of dark energy,
which has become a matter of intense debate \cite{debate}. By far
the simplest dark energy candidate is the cosmological constant,
$\Lambda$. However, albeit it fits reasonably well into the
cosmological data it faces two serious drawbacks on the
theoretical side. In the first place, its quantum field value
comes about 123 orders of magnitude larger than that observed.
Secondly, it gives rise to the {\em coincidence problem}: ``why
are the vacuum and dust energy densities of precisely the same
order today?" Bear in mind that the energy density of dust
red-shifts with expansion as $a^{-3}$, where $a$ denotes the scale
factor of the Friedmann--Robertson--Walker (FRW) metric. This is
why a number of candidates of various degrees of plausibility have
been proposed over the last few years with no clear winner in
sight. Here we focus on a dark energy candidate grounded on sound
thermodynamic considerations that is receiving growing attention
in the literature, namely, the ``holographic dark energy". It
arises from the holographic principle (which as formulated by 't
Hooft  and Susskind \cite{formulated} says that the number of
degrees of freedom of a physical system should scale with its
bounding area rather than with its volume) and the realization
that it should be constrained by infrared cutoff \cite{cohen}. On
these basis, Li \cite{Li} suggested the following
constraint on its energy density $\rho_{X} \leq 3 M_{P}^{2}%
c^{2}/L^{2} \,$ , the equality sign holding only when the
holographic bound is saturated. In this expression $M_{P}$ stands
for the reduced Planck mass, $c^{2}$ is a dimensionless constant
and $L$ denotes the infrared cutoff radius. The latter is not
specified at all by the holographic principle and different
options have been tried with different degrees of success, namely,
the particle horizon \cite{fischler}, the future event horizon
\cite{future}, and the Hubble horizon \cite{hsu,plb,sub}.

Scalar-tensor theories of gravity has been widely applied in
cosmology (see Faraoni's monograph \cite{valerio} for an
authorized review and Ref. \cite{gannouji} for a recent work) and
very  recently also in connection to holographic energy
\cite{setare}. The aim of this Letter is to build a cosmological
model of late acceleration based on the Brans-Dicke theory of
gravity \cite{b-d} and on the assumption that the (pressureless)
dark matter and holographic dark energy do not conserve separately
but interact with each other in a manner to be specified below. At
this point the interaction (coupling) may look purely
phenomenological but different Lagrangians have been proposed in
support of it -see \cite{lagrangians} and references therein. On
the other hand, in the absence of a symmetry that forbids the
interaction there is nothing, in principle, against it. Further,
the interacting dark mater-dark energy (the latter in the form of
a quintessence scalar field and the former as fermions whose mass
depends on the scalar field) has been investigated at one quantum
loop with the result that the coupling leaves the dark energy
potential stable if the former is of exponential type but it
renders it unstable otherwise \cite{dj}. So, microphysics seems to
allow enough room for the coupling; however, this point is not
fully settled and should be further investigated. The difficulty
lies, among other things, in that the very nature of both dark
energy and dark matter remains unknown whence the detailed form of
the coupling cannot be elucidated at this stage.

As infrared cutoff we shall choose the Hubble horizon -i.e., $L =
H^{-1}$- as it seems more natural. As it turns out, the transition
from decelerated to accelerated expansion is more readily effected
and the coincidence problem gets substantially alleviated, both
for spatially flat and curved FRW spaces. Our work differs from
that of Ref. \cite{setare} in many important respects, notably in
that we take the Hubble length as infrared cutoff not the future
event horizon, and that the author of \cite{setare} assumes that
the dark energy does not interact with matter. On the other hand,
while we focus on alleviating the coincidence problem and
providing a natural transition from decelerated to accelerated
expansion, the latter aims to obtain the conditions the model must
fulfill to prevent that the equation of state of the dark energy
crosses the phantom divide.

This Letter is outlined as follows. Section II presents the model.
Section III extends it to the non-flat case. Section IV examines
the implications of a  varying $M_{P}$. Lastly, Section V
summarizes our conclusions and provides some final remarks.

\section{The spatially flat FRW case}
For a spatially flat FRW universe filled with dust (dark matter)
and dark energy, the Brans--Dicke field equations take the form
\\
\begin{equation}
3\frac{{\dot{a}}^2}{a^2} = \frac{1}{{\phi} M_{P}^{2}}({\rho}_{M}+
{\rho}_X) + \frac{1}{2} \,{\omega}\, \frac{\dot{\phi}^2}{\phi ^2}
- 3\frac{\dot{a}\dot{\phi}}{a\phi}\; , \label{feq1}
\end{equation}
and
\begin{equation}
2\frac{\ddot{a}}{a} + \frac{{\dot{a}}^{2}}{a ^2} = -
\frac{1}{{\phi} M_{P}^{2}} \, p_{X} - \frac{1}{2}\, {\omega}\,
\frac{\dot{\phi}^2}{\phi ^2} - 2\frac{\dot{a}\dot{\phi}}{a\phi}
-\frac{\ddot{\phi}}{\phi}\; ,
 \label{feq2}
\end{equation}
\\
where $\omega$ stands for the Brans-Dicke parameter \cite{b-d}.

As stated above, both components -the pressureless dark matter and
the holographic dark energy- are assumed to interact with each
other; thus, one may grow at the expense of the other. The
conservation equations for them read
\\
\begin{equation}
\dot{\rho}_{M} + 3H{\rho}_{M} = Q , \quad \qquad \dot{\rho}_X +%
3H(1+ w) \rho_X= - Q \, , \label{conserv}
\end{equation}
\\
where $w \equiv p_{X}/\rho_{X}$ denotes the equation of state
parameter for the dark energy, and $Q$ stands for the interaction
term. Following \cite{plb,sub} we shall assume for the latter the
ansatz $ Q = \Gamma {\rho}_X$ with $\Gamma > 0$ being the
interaction rate which, in general, can vary with time.

The wave equation for the Brans-Dicke scalar field, $(2{\omega}%
+3)[\ddot{\phi} + 3 (\dot{a}/{a})\, \dot{\phi}] = T$ -where $T$
denotes the trace of the stress-energy tensor of dark matter and
dark energy- is not an independent expression as it follows from
the Bianchi identities alongside Eqs.
(\ref{feq1})-({\ref{conserv}). This wave equation is not altered
by the interaction (\ref{conserv}) since although the matter and
dark energy components do not conserve separately the overall
fluid -matter plus dark energy- does.

Taking up  Li's expression, with the equality sign, for  the
holographic dark energy \cite{Li} and  $L = H^{-1}$, we write
\\
\begin{equation}
{\rho}_X = 3c^2 M_{P}^{2} H^{2} \, .
 \label{rhox}
\end{equation}

At this point our system of equations is not closed and we still
have freedom to choose one. We shall assume that Brans-Dicke field
can be described as a power law of the scale factor, $\phi \propto%
a^{n}$. In principle there is no compelling reason for this
choice. However, we shall see in due course that for small $|n|$
it leads to consistent results. Thus, a partial justification will
be seen a posteriori.

By combining Eq. (\ref{rhox}) with the above expression for $\phi$
and the field equations (\ref{feq1}) and (\ref{feq2}), we get
\\
\begin{equation}
\dot{{\rho}}_{X} = - \frac{6c^{2}\, M_{P}^{2} H^{3}}{2+n} \left[1+
\frac{w}{(1+r)} \left(3(1+n) - \frac{n^{2} \,{\omega}}{2} \right)
- (n^{2}\, {\omega} +n^2 -n)\right] \, , \label{dotrx}
\end{equation}
\\
where $r \equiv \rho_{M}/\rho_{X}$. Inserting this into the second
equation of (\ref{conserv}), we obtain  an expression for the
equation of state parameter of the dark energy,
\\
\begin{equation}
w = (1+r)\, \frac{(3n^2 {\omega} +2n^2 - 5n) +
(n+2)\frac{\Gamma}{H}}{3[n -(n+2)r] -n^{2}\,{\omega}}. \label{w1}
\end{equation}
\\
It is important to note that if $n$ is zero, the Brans-Dicke
scalar field $\phi$ becomes trivial, and the last two equations
reduce to their respective expressions in general relativity
\cite{plb}. Equation (\ref{w1}) clearly shows that $w$ can be
negative. This requirement only puts some bounds on the values of
$n$ and ${\omega}$. A case of particular interest is that when
$|n|$ is small whereas ${\omega}$ is high so that the product
$n^2{\omega}$ results of  order unity. This is interesting because
local astronomical experiments set a very high lower bound on
${\omega}$ \cite{will1,will2}; in particular, the Cassini
experiment \cite{bertotti} implies that $\omega > 10^{4}$.
Likewise, a slow fractional variation of $\phi$ will lead to a
small fractional variation of $G$, consistent with observations.
In this case $w$ takes a simpler form,
\\
\begin{equation}
w \simeq - \frac{(1+r)}{6r + n^2 {\omega}}\left(3n^2 {\omega} +
2\frac{\Gamma}{H}\right). \label{w2}
\end{equation}
\\
It is clearly negative-definite and decreases with expansion
whenever $\Gamma/H$ augments with time.

Now, as the dynamics of the scale factor is governed not only by
the dark matter and the holographic dark energy, but also by the
Brans-Dicke field, the signature of the deceleration parameter, $q %
= -\ddot{a}/(aH^{2})$, has to be examined carefully. If we divide
Eq. (\ref{feq2}) by $H^2$, combine the resulting expression with
Eq. (\ref{rhox}) and the relationship $\phi \propto a^{n}$, we get
\\
\begin{equation}
q = \frac{3wc^2}{(2+n)\phi} + \frac{n^2{\omega}}{2(2+n)} +
\frac{n^2 + n + 1}{2+n}. \label{q1}
\end{equation}
\\
So, with a negative $w$, $q$ can obviously be negative if
\\
\[
\left|\frac{3wc^2}{(2+n)\phi}\right| > \left|\frac{n^2 +n
+1}{2+n}\right|+ \left|\frac{n^2 \omega}{2(2+n)}\right|.
\]

It is interesting to note that although $q$ does not contain
$\Gamma$ or $r$ explicitly, it actually depends on these two via
the expression for $w$.  If $\phi$ decreases with $a$, i.e., if
$n<0$, then the absolute value of the first term in Eq. (\ref{q1})
will increase and $q$ might also have a signature flip from a
positive to a negative value. With $|n| <<1$ and $
(n^{2}{\omega})\sim {\cal O}\, (1)$, this equation reads as
\\
\begin{equation}
q \simeq \frac{3wc^2}{2\phi} + \frac{n^2{\omega}}{4} +
\frac{1}{2}\, .
\label{q2}
\end{equation}

This implies a clear improvement with respect to the model of Ref.
\cite{plb}. There, no acceleration can be achieved in the absence
of interaction. Here, even in the non-interacting limit ($\Gamma = %
0$) we can have $q < 0$ -see Eqs. (\ref{w2}) and (\ref{q1}).
Besides, to stage a transition from deceleration to acceleration
in \cite{plb} it was necessary that the quantity $c^2$, entering
the expression for the holographic dark energy density (Eq.
(\ref{rhox})), should be slightly dynamical. While, that
assumption seems to us very reasonable as there is no reason why
the holographic bound should be already saturated at present, in
the model considered in this Letter it is not necessary at all,
though a slowly varying $c^2$ may also help the transition.

Our holographic interacting model shares with the model of Ref.
\cite{plb} the  advantage of considerably alleviating the
coincidence problem. It is alleviated in the sense that the ratio
between the energy densities of matter and dark energy, $r$, can
vary more slowly in this model than in the conventional
$\Lambda$CDM model, where $|\dot{r}/r|_{0} = 3H_{0}$ (here and
throughout  a zero subscript indicates present time). Indeed, by
virtue of the conservation equations ({\ref{conserv}) the
evolution for the aforesaid ratio is seen to obey
\begin{equation}
\dot{r}= 3Hr \left[w + \frac{1+r}{r} \frac{\Gamma}{3H}\right]\, .
\label{dotr}
\end{equation}
\\
Therefore, keeping in mind that at present $w\simeq{-1}$
\cite{wmap3,seljak}, a ``soft" coincidence can be achieved if $0 < %
(\Gamma/3H)_{0} < (1+r_{0})/r_{0}$. This is consistent with the
requirement that $\Gamma > 0$. It should be noted that the field
equations (\ref{feq1}) and (\ref{feq2}) do not enter the
derivation of (\ref{dotr}). Thus, the latter remains the same as
that in general relativity -see Eq. (5) of Ref. \cite{plb}.

\section{Curved FRW cases}
As is widely believed, inflation practically washes out the effect
of curvature in the early stages of cosmic evolution. However, it
does not necessarily imply that the curvature has to be wholly
neglected at  present. Indeed, aside from the sake of generality,
there are sound reasons to include it: $(i)$ Inflation drives the
$k/a^{2}$ ratio close to zero but it cannot set it to zero if $k
\neq 0$ initially. $(ii)$ The closeness to perfect flatness
depends on the number of $e$-folds and  we can only speculate
about the latter. $(iii)$ After inflation the absolute value of
the $k/a^{2}$ term in the field equations may increase with
respect to the matter density term, thereby the former should not
be ignored when studying the late Universe. $(iv)$ Observationally
there is room for a small but non-negligible spatial curvature
\cite{wmap3,seljak}.

After incorporating the curvature term, the field equations
(\ref{feq1}) and (\ref{feq2}) generalize to
\begin{equation}
3\frac{{\dot{a}}^2}{a^2} +3\frac{k}{a^2} =\frac{1}{{\phi}
M_{P}^2}({\rho}_{M}+ {\rho}_X) + \frac{1}{2} {\omega}\,
\frac{\dot{\phi}^2}{\phi ^2} - 3\frac{\dot{a}\dot{\phi}}{a\phi}\,
, \label{feqk1}
\end{equation}
\\
and
\\
\begin{equation}
2\frac{\ddot{a}}{a} + \frac{{\dot{a}}^{2}}{a ^2}+ \frac{k}{a^2}= -
\frac{1}{{\phi} M_{P}^2}\, p_{X} - \frac{1}{2} {\omega}\,
\frac{\dot{\phi}^2}{\phi ^2} - 2\frac{\dot{a}\dot{\phi}}{a\phi}
-\frac{\ddot{\phi}}{\phi} \, ,
\label{feqk2}
\end{equation}
\\
respectively. The equation of state parameter of dark energy -Eq.
(\ref{w1})- now reads
\\
\begin{equation}
w = (1+r) \, \frac{(3n^2 {\omega} +2n^2 - 5n) +
(n+2)\frac{\Gamma}{H}+ 10 \, \Omega_{k}}{3[n(1-r) -2r]-n^{2}
{\omega}-6 \, \Omega_k} \, , \label{w3}
\end{equation}
where $\Omega_{k} \equiv -k/(a^{2}H^{2})$.

The deceleration parameter  follows from Eq. (\ref{feqk2}) and
takes the form
\\
\begin{equation}
q = \frac{3wc^2}{(2+n)\phi} + \frac{n^2{\omega}}{2(2+n)} +
\frac{n^2 + n + 1}{2+n} - \frac{\Omega_{k}}{(2+n)}\, \, .
\label{q3}
\end{equation}

Thus, the spatial curvature does not seriously modify the
qualitative picture of the previous section but it may however
affect the time of onset of the acceleration. For an open universe
$(\Omega_{k}>0)$ the acceleration sets in earlier whereas in a
closed universe $(\Omega_{k} <0)$ the accelerated phase is
delayed. A more careful look at the equation for $w$ reveals that
in the second case, i.e., for a closed universe, a further
condition has to be satisfied. For example, in the small $|n|$
limit, $3n^{2} \omega + \frac{2\Gamma}{H} + 10 \, \Omega_{k}> 0$,
together with $6r + n^{2} \omega + 6 \, \Omega_{k} > 0$, or the
reversed direction of the inequality for both. For an open
universe, however, there is no further restriction on the onset of
acceleration.

\section{A model with a varying $M_{P}$}
It is well known that as a result of the non-minimal coupling
between the scalar field and the Ricci scalar, in Brans-Dicke
theory, the gravity ``constant" $G$ is no longer a constant. The
relation between $G$ and $\phi$ is given by $G = G_{N}/\phi$ with
$G_{N}$ denoting its Newtonian (constant) value. The reduced
Planck mass is defined as
\\
\begin{equation}
M_{P}^{2} = \frac{1}{8\pi\, G}\, . \label{mp}
\end{equation}

In the previous section it was tacitly assumed that $M_P$ was a
constant given by $M_{P}^{2} = 1/(8\pi\, G_{N})$. One may be
tempted to check what happens in this case if we stick to the
relation (\ref{mp}). Here, the behavior of a spatially flat FRW
model $(k=0)$ is considered. As equation (\ref{mp}) yields that
$M_{P}^{2} \, \phi$ is a constant, the equation for the
deceleration parameter becomes independent of $\phi$, i.e.,
equation (\ref{q1}) now reads
\\
\begin{equation}
q = \frac{3wc^2}{n+2} + \frac{n^2{\omega}}{2(n+2)} +\frac{n^{2}
+n+1}{n+2}\, ,
 \label{q4}
\end{equation}
\\
whereas the expression for the equation of state parameter becomes
\\
\begin{equation}
w = (1+r) \frac{3(n^2 {\omega} +n^2 -n) +
(n+2)\frac{\Gamma}{H}}{3[n(1-r) -2r] -n^2 {\omega}}.
\label{w4}
\end{equation}
\\
So, there is hardly any change in the qualitative behavior of the
scenario. In fact, when $|n| \ll 1$ it can be checked that Eq.
(\ref{w4}) reduces to (\ref{w2}). The time behavior of $q$, more
specifically the possibility of a signature flip, has now to be
taken care of by the variation of $w$.

\section{Discussion}
As the accelerated expansion of the Universe seems to be a
comparatively recent episode and must have taken over from the
more sedate decelerated expansion in the matter dominated era
itself, it is good to have a signature flip in the deceleration
parameter $q$ from a positive to a negative value in this era.
This holographic dark energy model in Brans-Dicke theory serves
this purpose even with a constant $c^{2}$ whereas the same kind of
a dark energy yields an ever-accelerated expansion in general
relativity \cite{plb}.

It is important to note that Brans-Dicke theory, either by itself
or together with some form of dark energy, indeed provides a model
to the accelerated expansion. However, these are all with a very
small value of ${\omega}$ \cite{small}. But it is quite well known
now that local astronomical experiments  severely restricts this
parameter to very high values \cite{bertotti}. One important
feature of the present model is that with a small value of $|n|$,
${\omega}$ can indeed have a high value.

The fractional rate of change of $G$, $|\dot{G}/G| = |nH|$, which
follows from our simplifying assumption that $\phi \propto a^{n}$
with a low value of $|n|$, is quite consistent with the
observational requirement that $|\dot{G}/G|_{0}$ must be lower
than the current value of Hubble's expansion rate \cite{jofre}
-for a long list of experiments, see \cite{vitaly}. (In
retrospect, this could serve as a motivation for the assumption).

There is another important point. In this model, the interaction
between the dark energy and the dark matter is certainly crucial.
It is interesting to note that while in a similar  general
relativity model, \cite{plb}, there is no non-interacting limit
which yields an acceleration, in Brans-Dicke theory indeed such a
possibility does exist. Equations (\ref{w2}) and (\ref{q1})
readily show that $q$ can have negative values even for vanishing
$\Gamma$.

As is well known, holographic energy is not compatible with
phantom energy \cite{no-phantom} thereby we must impose $w \geq
-1$. This, combined with Eq. (\ref{w2}), yields
\[
r \geq \frac{2 \left( n^{2}\, \omega + \frac{\Gamma}{H}
\right)}{6- 3n^{2} \omega - \frac{2\Gamma}{H}}.
\]
\\
This constraint is certainly satisfied provided both $n^{2}\,
\omega$ and $ \Gamma/H$ do not exceed order one.

Clearly, it should be highly desirable to be in position to
determine $\Gamma$ from a fundamental theory but, as hinted in the
Introduction, we are far from it. Nevertheless, observational
astrophysics might soon be able  to set reliable bounds on the
present value of $\Gamma$. In this connection, Tetradis {\it et
al.} \cite{tetradis} have devised ways based on the rates of
direct detection of dark matter particle that live in the Galaxy
halo to establish limits on $\kappa$, the ratio between the
strength of the dark matter-dark energy interaction and the
strength of Newtonian gravity. In particular, they found that the
average and maximum energy of the dark matter particles varies as
$1+\kappa^{2}$. However, before their method may be applied
unambiguous direct detection of the particles is mandatory.

Associated to a never-ending accelerated era there is a future
event horizon, the existence of which means a serious handicap for
string theories. This is why it has been speculated that the
current phase of accelerated expansion must be followed by a fresh
decelerated era and cosmological models featuring such transition
back to a decelerated expansion has been proposed -see, e.g.,
\cite{fresh}. Inspection of the three pair of expressions for $w$
and $q$ reveals that, in each case,  our model can achieve such a
transition provided the interaction rate, $\Gamma$, evolves to
negative values. Clearly, such a possibility looks contrived.
However, we should not wonder since the proposal of reverting to a
decelerated era just for the purpose of getting rid of the event
horizon appears unnatural, especially because no observational
data suggests it. Nevertheless, this possibility cannot be
discarded right away whereby we should keep an open mind. In any
case, we would like to remark that holographic dark energy models
that identify the infrared cutoff $L$ with the future event
horizon cannot account for such a transition.

It is also noteworthy that if at present $-1<w<-1/3$ and the
interaction rate is not high compared to $H_{0}$, i.e.,
$(\Gamma/3H)_{0}$ is small, then it is quite possible to have
$|\dot{r}/r|_{0} <H_{0}$. Put in another way, the model
significantly alleviates the coincidence problem.

\acknowledgments{We are grateful to Winfried Zimdahl for comments
and suggestions on an earlier draft of this Letter. N.B. was
funded by the UAB-CIRIT Grant 240252. This work was partly
supported by the Spanish ``Ministerio de Educaci\'{o}n y Ciencia"
under Grant FIS2006-12296-C02-01.}

\end{document}